\documentclass[prb,aps,twocolumn,showpacs,superscriptaddress]{revtex4-2}
\usepackage{amsfonts,amsmath,multirow}
\usepackage[T1]{fontenc}
\usepackage[utf8]{inputenc}

\usepackage{graphicx}
\usepackage{xcolor}
\usepackage{epstopdf}
\usepackage{dsfont}
\usepackage{physics}
\usepackage{hyperref}
\usepackage{lipsum}
\usepackage{dcolumn}
\usepackage{orcidlink}
\usepackage{natbib}

\usepackage{mathtools}
\usepackage{hf-tikz}
\usepackage[normalem]{ulem}

\usepackage[]{newtxtext} %arxiv OFF
\usepackage[nosymbolsc,smallerops,bigdelims]{newtxmath} %arxiv OFF
\DeclareMathAlphabet{\mathcal}{OMS}{cmsy}{m}{n} %arxiv OFF
\DeclareMathAlphabet{\mathbcal}{OMS}{cmsy}{b}{n} %arxiv OFF

\usepackage{bm}

\newcommand{\kp}{k\! \vdot\! p}

\newcommand{\rr}{\bm{r}}

\DeclareUnicodeCharacter{0141}{\L{}}

\definecolor{bred}{HTML}{e31a1c}
\definecolor{bgreen}{HTML}{33a02c}
\definecolor{bblue}{HTML}{1f78b4}

\definecolor{armygreen}{rgb}{0.29, 0.33, 0.13}
\definecolor{newred}{RGB}{255,70,70}
\definecolor{newcyan}{RGB}{0,200,255}

\newcolumntype{L}{D{.}{.}{3,4}}

\begin{document}
	
\title {Light-hole states and hyperfine interaction in electrically-defined Ge/GeSn quantum dots}

    \author{Agnieszka Miętkiewicz}
    \affiliation{Institute of Theoretical Physics, Wroc\l aw  University of Science and Technology, Wybrze\.ze Wyspia\'nskiego 27, 50-370 Wroc{\l}aw, Poland}

    \author{Jakub Ziembicki\orcidlink{0000-0002-6332-6986}}
    \affiliation{Department of Semiconductor Materials Engineering, Wroc\l aw  University of Science and Technology, Wybrze\.ze Wyspia\'nskiego 27, 50-370 Wroc{\l}aw, Poland}

    \author{Krzysztof Gawarecki\orcidlink{0000-0001-7400-2197}}
    \email{Krzysztof.Gawarecki@pwr.edu.pl}
    \affiliation{Institute of Theoretical Physics, Wroc\l aw  University of Science and Technology, Wybrze\.ze Wyspia\'nskiego 27, 50-370 Wroc{\l}aw, Poland}

    \begin{abstract}
         We theoretically investigate hole spins confined in a gate-defined quantum dot (QD) embedded in GeSn/Ge/GeSn quantum well (QW) structure. Owing to the tensile strain in the Ge layer, the system effectively realizes a light-hole qubit. We systematically study how various morphological parameters influence the energy spectrum and the hyperfine coupling to the nuclear spin bath. The simulations are carried out using a realistic, fully atomic sp$^3$d$^5$s$^*$ tight-binding model. We also perform complementary DFT calculations of wave functions near the atomic cores and use them to parameterize the hyperfine‐interaction Hamiltonian. We evaluate the Overhauser field fluctuations and demonstrate that the strength of the hyperfine coupling for the lowest hole doublet crucially depends on the Sn content in the barrier. We highlight the conduction-valence band mixing, which leads to considerable $s$-type admixtures to the hole states, providing the dominant channel of hyperfine coupling due to the Fermi contact interaction.
         
\end{abstract}
	
	\maketitle
	
\section{Introduction}
\label{sec:intr}

Hole-spin qubits are among the most promising candidates for quantum information processing due to their compatibility with all-electrical control schemes and long coherence times~\cite{Terrazos2021,Fang2023}. The group IV material systems provide high CMOS integrability, a key requirement for scalable, application-driven architectures, while also offering lower toxicity and substantially reduced material costs compared with III–V compounds. The hole spin states in electrically-defined germanium QDs gained much attention due to a high degree of tunability~\cite{Hendrickx2020,Bosco2021}. Also, owing to a low natural abundance of Ge stable isotopes carrying non-zero nuclear spin, the hyperfine interaction is much weaker compared to InAs and GaAs systems, leading to an enhanced coherence time~\cite{Fang2023}.

In most germanium-based nanostructures, such as Ge/SiGe quantum wells (QWs), the Ge layer is compressively strained in-plane, which leads to the heavy-hole (HH) character of the hole ground state. However, the opposite scenario with the ground light-hole (LH) states is beneficial for quantum technological applications~\cite{Assali2022,Huo2014}, such as coherent photon polarization to electron spin conversion~\cite{Vrijen2001}. Recently, the Ge/GeSn systems gained attention~\cite{Assali2022,DelVecchio2023,DelVecchio2024,DelVecchio2025}, where the Ge QW experiences tensile-strain due to a larger Sn lattice constant compared to Ge, which leads to the desired LH ground-state scenario. This was also proposed as a realization of the gated quantum dot, where the lateral electric field provides in-plane confinement in GeSn/Ge/GeSn QW~\cite{DelVecchio2023}. The electric and spin-related properties of such QDs were investigated in terms of the $\kp$ approach~\cite{DelVecchio2023}. It has been shown that the LH qubit exhibits a significantly larger electric dipole moment than the HH states, enabling efficient electric dipole spin resonance (EDSR) control.

In planar Ge/GeSn heterostructures, the large linear-in-$k$ Rashba spin-orbit interaction (RSOI) intrinsic to light-hole states is highly tunable via out-of-plane electric fields~\cite{DelVecchio2025}. These systems exhibit unique spin-switching capabilities, where RSOI vanishes at specific gate voltages, enabling a fully electrical control between ``on'' and ``idle'' spin qubit modes~\cite{Bosco2021,DelVecchio2025}. These characteristics, combined with the scalability and LH ground state enabled by tensile strain from GeSn substrates, make Ge/GeSn quantum dots especially promising platforms for scalable quantum processors~\cite{DelVecchio2025,DelVecchio2023}. Additionally, these systems show improved resilience to charge noise, which is an important factor affecting coherence times~\cite{DelVecchio2023, Bosco2021}.

Tensile strain in Ge within Ge/GeSn systems turns Ge into a direct bandgap semiconductor when the strain exceeds 1.8\%~\cite{DelVecchio2023}, which is essential for efficient photon--spin qubit interfaces~\cite{DelVecchio2023,DelVecchio2024}. In terms of decoherence, the hyperfine interaction (HF) with surrounding nuclear spins is a major limiting factor for electron spins~\cite{Khaetskii2002,Fischer2008,Schliemann2003}. However, the hole--nucleus coupling is significantly weaker than the electron--nucleus interaction due to the (dominant) $p$-orbital symmetry of the hole wavefunction~\cite{Bulaev2007,Machnikowski2019}. Moreover, the natural abundance of the $^{73}$Ge isotope --- the only stable Ge isotope with non-zero nuclear spin --- is relatively low (about 7.76\%)~\cite{Schliemann2003}. Finally, a thorough understanding of hyperfine coupling mechanisms enables more sophisticated approaches than merely suppressing the interaction, such as storing quantum information in nuclear spin states~\cite{Fang2023, Taylor2003}.

In this paper, we study theoretically the energy levels and the hyperfine interaction for a light-hole qubit in GeSn/Ge/GeSn quantum dot. Within the sp$^3$d$^5$s$^*$ tight-binding model, we simulate the behavior of the ground doublet of states under various conditions. We also perform the state-of-the-art DFT calculations to find the parameters describing the hyperfine interaction.
We calculate the fluctuations of the Overhauser field, and analyze the impact of various coupling channels. We find that in the broad range of system parameters, the hyperfine interaction is dominated by the $s$-type contributions to the light-hole wave function, which allow the contact interaction. Our results demonstrate that an accurate description of Ge/GeSn QD systems is possible only within the models that capture the conduction-valence band mixing effects.

The paper is organized as follows. In Section~\ref{sec:model}, we introduce the models for the system geometry, the hole states, and the hyperfine interaction. The subsequent Section~\ref{sec:results} contains the numerical results and the discussion. The Section~\ref{sec:concl} concludes the paper. Finally, in the Appendix, we provide details on the strain modeling, the DFT wave functions, and the Overhauser‐field fluctuation calculations.

\section{Model}
\label{sec:model}
\subsection{Structure}

\begin{figure}
 \begin{center}
     \includegraphics[width=3.5in]{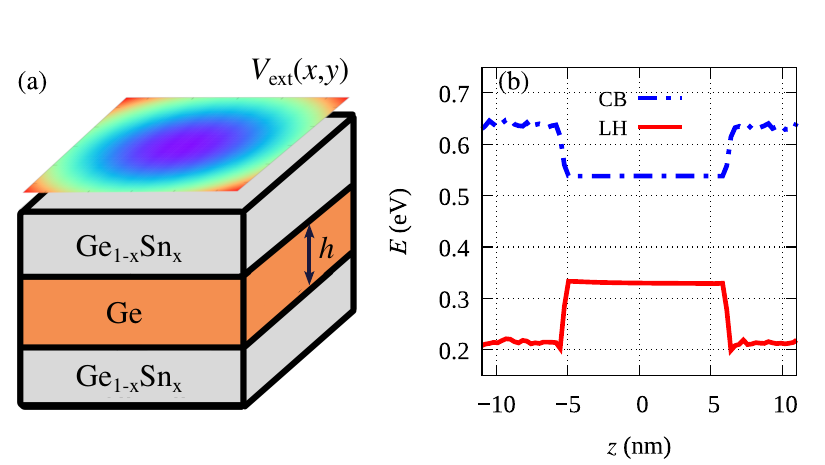}
     \end{center}
     \caption{\label{fig:schema} (a) Schematic picture of the material layers in the considered system, and the quadratic potential $V_\mathrm{ext}(x,y)$ creating the in-plane confinement; and (b) the $\Gamma$-point band edges for Ge$_{0.85}$Sn$_{0.15}$ in the barrier.}
\end{figure}

The system under consideration is built on a Ge quantum well of thickness $h$ embedded in Ge$_{1-x}$Sn$_x$ barrier, as depicted in Fig.~\ref{fig:schema}(a). Tensile strain in the QW layer, which is caused by the lattice mismatch between Sn and Ge, shifts the LH subband upward with respect to the valence-band edge in the barrier, leading to hole confinement~\cite{Assali2022}. The band edges at the $\Gamma$ point of the Brillouin zone (BZ) are shown in Fig.~\ref{fig:schema}(b). These band‐edge energies were computed by constructing virtual bulk model matching the local composition and strain, performing tight-binding calculations, and computing the average over atoms in the plane.

The quantum dot is formed in the Ge layer by applying an external electric field, creating an in-plane potential $V_\mathrm{ext}(x,y)$, as schematically shown in Fig.~\ref{fig:schema}(a). Therefore, the system can be described by
\begin{align}
\label{eq:ham}
    H = H_0 + V_\mathrm{ext}(x,y) ,
\end{align}
where $H_0$ is the tight-binding Hamiltonian for the GeSn/Ge/GeSn structure. % and $\mathbb{I}$ is the unit operator. 
The gate-defined potential $V_\mathrm{ext}(x,y)$ is assumed to have the form
\begin{equation*}
    V_\mathrm{ext}(x,y) = \frac{m^*_\mathrm{lh} \omega^2}{2} \qty(x^2 + y^2),
\end{equation*}
where $m^*_\mathrm{lh}$ is the light-hole effective mass. This potential is attractive for holes but repulsive for electrons.

\subsection{The hole states}
\label{sec:states}
We calculate the hole states within the atomistic sp$^3$d$^5$s$^*$ tight-binding model~\cite{Slater1954, Jancu1998, Zielinski2010,Gawarecki2025}. Since the model covers the full BZ, it is well-suited for studies on GeSn systems, which experience the indirect-direct band-gap transition~\cite{Gawarecki2024}. The Hamiltonian at zero electric field can be written as
\begin{align*}
H_0 &= \sum^{N_\mathrm{A}}_{i} \Bigg \{  \sum_{\alpha} E^{(i)}_{\alpha} a^\dagger_{i,\alpha} a_{i,\alpha} +  \sum^{N_\mathrm{A}}_{j\neq i} \sum_{\alpha,\beta} t^{(ij)}_{\alpha\beta} a^\dagger_{i,\alpha} a_{j,\beta}  \\ & \phantom{=} +  \sum_{\alpha, \beta} \Delta^{(i)}_{\alpha \beta} a^\dagger_{i,\alpha} a_{i,\beta} \Bigg \} ,
\end{align*}
where $N_\mathrm{A}$ is the total number of atoms in the system, $E^{(i)}_{\alpha}$ are the on-site orbital energies, $t^{(ij)}_{\alpha\beta}$ are hopping integrals, and $\Delta^{(i)}_{\alpha \beta}$ accounts for the spin-orbit coupling. The material parameters and implementation details are given in Ref.~\cite{Gawarecki2024}.

With the position matrix elements approximated by $\mel{\bm{R}_i; \alpha}{x_n}{\bm{R}_j; \beta} \approx (\bm{R}_i)_n \, \delta_{i j} \, \delta_{\alpha \beta}$, the electric-field potential is written in a second quantization form
\begin{align*}
V_\mathrm{ext} &=  \frac{m^*_\mathrm{lh} \omega^2}{2} \sum^{N_\mathrm{A}}_{i} \sum_{\alpha} \qty[(\bm{R}_i)^2_x + (\bm{R}_i)^2_y]  a^\dagger_{i,\alpha} a_{i,\alpha},
\end{align*}
where we took the effective mass $m^*_\mathrm{lh} = 0.043 m_0$~\cite{ioffe_Ge}, which is the bulk value for LH in Ge. The magnetic field is incorporated into the model via Peierls substitution~\cite{Graf1995,Boykin2001,Vogl2002}, as well as the orbital- and spin Zeeman on-site matrix elements~\cite{Ma2016,Gawarecki2025}.

The strain is taken into account by calculating the atomic displacements through minimization of the elastic energy of the system. To this end, we utilize the Martin's Valence Force Field (VFF) model~\cite{Martin1970,Tanner2019}, as described in Ref.~\cite{Gawarecki2025}. The model parameters for Ge and Sn are extracted from the elastic constants $C_{11}$, $C_{12}$, $C_{44}$, and the Kleinmann parameter, using analytical relations~\cite{Tanner2019}. The resulting values are listed in Table~\ref{tab:elastic} in Appendix~\ref{app:strain}.

To facilitate further calculations of the hyperfine interaction, we use the basis of the orbital angular momentum eigenstates, i.e. \{$s$, $p_{-1}$, $p_{0}$, $p_{1}$, $d_{-2}$, $d_{-1}$, $d_{0}$, $d_{1}$, $d_{2}$, $s^*$\} with the spin configurations $\uparrow / \downarrow$ (20 basis states in total). 

\subsection{The hyperfine interaction}

\begin{table}
    \caption{\label{tab:hf} The parameters used in the hyperfine interaction calculations. The values of $I$, $\mu_I$, and the natural abundance are taken from Ref.~\onlinecite{Schliemann2003}, while $\zeta$ are obtained from $\mu_I = \zeta \mu_N I$. The values of $\mathcal{R}_S(0)$ and  $M_\mathrm{\alpha\beta}$ are calculated using the DFT wave functions.}
    \begin{ruledtabular}
        \begin{tabular}{lcccc}
            & \hspace{20pt} Ge$^{73}$ \hspace{20pt} & Sn$^{115}$ \hspace{10pt} & \hspace{5pt} Sn$^{117}$ \hspace{10pt} & \hfill Sn$^{119}$   \\
            \hline \\[-0.5em]
            $I$  & 9/2 & \hspace{-18pt}1/2 & \hspace{-12pt}1/2 & 1/2   \\[1.1pt]
            $\mu_I$  & -0.8792 &\hspace{-18pt} -0.918 & \hspace{-12pt}-1.000 & -1.046 \\[1.1pt]
            $\zeta$  & -0.1954 & \hspace{-18pt}-1.836 & \hspace{-12pt}-2.000 & -2.092 \\[1.3pt]
            abundance \hfill(\%) & 7.76 & \hspace{-18pt}0.35 & \hspace{-12pt}7.16 & 8.58  \\[0.7pt]
            \hline \\[-0.5em]
            $\abs{\mathcal{R}_S(0)}^2$ \hfill(\AA$^{-1/3}$) & 2825 &  \multicolumn{3}{c}{5172} \\[1.1pt]
            $M_\mathrm{pp}$  &0.0369 & \multicolumn{3}{c}{0.0214}  \\[1.1pt]
            $M_\mathrm{dd}$  &0.0112 & \multicolumn{3}{c}{0.0043}  \\[1.1pt]
            $M_\mathrm{sd}$  &0.0004 & \multicolumn{3}{c}{0.0002}  \\[1.1pt]
        \end{tabular}
    \end{ruledtabular}
\end{table}

To quantify the interaction between the hole spin and the nuclei spins, we calculate the fluctuations of the Overhauser field for the lowest Zeeman doublet. For this purpose, we follow the implementation described in Refs.~\cite{Machnikowski2019,Gawarecki2025}. The hyperfine interaction Hamiltonian is written in the basis of the two lowest hole spin states
\begin{align*}
 H_{\mathrm{hf}}  = \frac{1}{2} \bm{h} \cdot \bm{\sigma},
\end{align*}
where $\bm{\sigma}$ is the vector of the Pauli matrices, %(related to the two-dimensional subspace of the lowest Zeeman doublet), 
and $\bm{h}$ is the Overhauser field. The fluctuations of the field components are represented by the root mean squares (rms), where we assume an unpolarized thermal state of the nuclei~\cite{Fischer2008,Machnikowski2019}
\begin{align}
    \label{eq:hn2}
    \expval{h^2_n} &= \frac{1}{3} \mu^2_N \sum_{i, m} I_i ( I_i + 1)  \qty(\zeta_i \Tr{A_m(\bm{r} - \bm{R}_i) {\sigma}_n} )^2,
\end{align}
where the index $i$ goes over all atomic sites; $n$ and $m$ are $x$, $y$, $z$ components; $\mu_N$ is the nuclear magneton, $I_i$ is the nuclear spin for $i$-th site, $\zeta_i$ is a parameter connecting the nuclear spin and the magnetic moment ($\bm{\mu}_I = \zeta \mu_N \bm{I}$),
\begin{equation*}
    \bm{A}(\bm{r})  = \frac{\mu_0 \mu_B}{2 \pi \hbar} \qty( \frac{8\pi}{3} \delta(\rr) \bm{S} + \frac{\bm{L}}{r^3} + \frac{3 (\hat{\bm{r}} \cdot \bm{S}) \hat{\bm{r}} - \bm{S}}{r^3}  ),
\end{equation*}
where $\mu_0$ is the vacuum permeability and $\mu_B$ is the Bohr magneton; $\bm{S}$ and $\bm{L}$ are the spin and orbital angular momentum operators, respectively. One should note that $\bm{A}$ contains three distinct terms: the contact part (which is nonvanishing only at the center of the nuclei), the orbital term, and the dipole term~\cite{Khaetskii2002,Testelin2009,Chekhovich2013}.
%, that describe the interaction of the carrier spin with the nuclei. 
Evaluating Eq.~\ref{eq:hn2} requires the matrix elements of $\bm{A}$ involving various TB orbitals. This, in turn, depends on the value of the $s$-type orbital at the center of the nuclei [the value of the radial function $\mathcal{R}_S(0)$], and the matrix elements 
\begin{equation*}
M_{\alpha \beta} = \frac{1}{\abs{\mathcal{R}_S(0)}^2} \int r^2 \mathcal{R}^*_\alpha(r) \frac{1}{r^3} \mathcal{R}_{\beta}(r)  \dd{r}.    
\end{equation*}
The tight-binding implementation is described in much detail in Ref.~\cite{Gawarecki2025}, and more briefly in Appendix~\ref{app:hf}. However, here in contrast to Refs.~\cite{Machnikowski2019,Gawarecki2025}, we do not use the hydrogen-like atomic wave functions. Instead, we derive the parameters based on wave functions obtained from DFT. 
%as described in the Appendix~\ref{app:dft}.

\subsection{DFT calculations}
\label{sec:dft}
To derive the values of $\mathcal{R}_S(0)$ and $M_{\alpha \beta}$, we performed DFT calculations for the bulk Ge and Sn using the VASP code \cite{VASP,VASP2,VASP3}. 
For an accurate description of these crystals, we used the LDA functional~\cite{LDA1,LDA2} for geometry optimization, and the hybrid functional~\cite{Hybrid} for the subsequent wave function calculations. Hartree-Fock exchange fraction was set to 0.19, which yields an accurate direct band gap for Ge (0.911 eV). The energy cut-off was set to 400~eV, and the $8\times8\times8$ $k$-point grid was used for the BZ sampling.

\begin{figure}[t]
    \centering
    \includegraphics[width=0.48\textwidth]{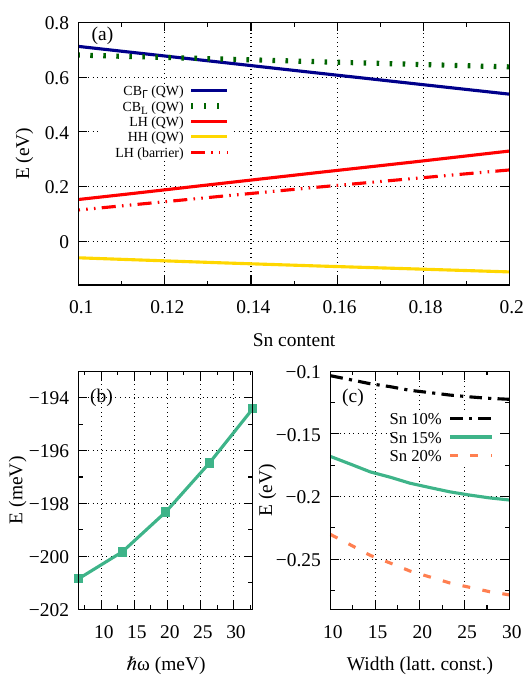}
     \caption{\label{fig:en_h} (a) The band edges at the $\Gamma$-point inside the QW (solid lines) and for the barrier (dashed line, only LH shown here), as a function of the barrier composition. The dotted line denotes the QW conduction-band edge at the $L$ point of the BZ. Zero energy corresponds to the unstrained Ge VB edge. (b) The dependence of the QD hole ground-state energy on $\hbar \omega$ for the fixed QW width. (c) Energy of the hole ground state in QD as a function of the QW thickness. } 
 \end{figure}

The VASP package employs the PAW method \cite{VASP_PAW} to represent atomic cores, which is accurate and efficient in the calculations for crystals. This implies that the all-electron (AE) wave function $\ket{\Psi_n}$ is replaced by a pseudo- (PS) wave function $\ket{\smash{\tilde{\Psi}_n}}$ with soft nodeless radial character near the atomic core. Therefore, we reproduced the real wavefunction on radial grid by following its definition in PAW method
\begin{equation*}
\ket{\Psi_n} = 
\ket{\smash{\tilde{\Psi}_n}} +
\sum_i \left( \ket{\phi_i} - \ket{\smash{\tilde{\phi}_i}} \right)
\langle \smash{\tilde{p}_i} | \smash{\tilde{\Psi}_n} \rangle,
\end{equation*}
where $\ket{\phi_i}$, $\ket{\smash{\tilde{\phi}_i}}$, $\ket{\tilde{p}_i}$ are AE partial waves, PS partial waves, and projector functions, respectively. This approach should, in principle, give the correct AE wave function in the region near the atomic core, which is relevant here, since $M_{\alpha \beta}$ depends primarily on the wave function in this region. The resulting values of the parameters for the hyperfine interaction are given in Table.~\ref{tab:hf}. More details on the DFT calculations can be found in Appendix~\ref{app:dft_wf}.

\section{Results}
\label{sec:results}

\subsection{Hole energy levels}

To establish a baseline for our study, we first characterized the GeSn/Ge/GeSn quantum well (i.e. we took $\omega = 0$). The conduction and valence band edges calculated as a function of the barrier composition Ge$_{1-x}$Sn$_x$ are shown in Fig.~\ref{fig:en_h}(a). As expected, the increasing tensile strain in the Ge layer, which is accompanied by the increasing Sn content in the barrier, results in the reduction of the effective band gap~\cite{Niquet2009}. 
The increase in the Sn content of the barrier leads to a larger offset between the energies in the QW and in the barrier, which gives stronger confinement in the QW layer. As expected, there is also a transition from the indirect to direct band gap regime~\cite{Zheng2018,Niquet2009,Polak2017}, which is at about 12.5\% of Sn in the barrier. This is reasonably close to the value of about $10.5$\% Sn, predicted by the $\kp$ model~\cite{DelVecchio2024}, yet with different parameters set.

\begin{figure}[t]
    \centering
    \includegraphics[width=0.48\textwidth]{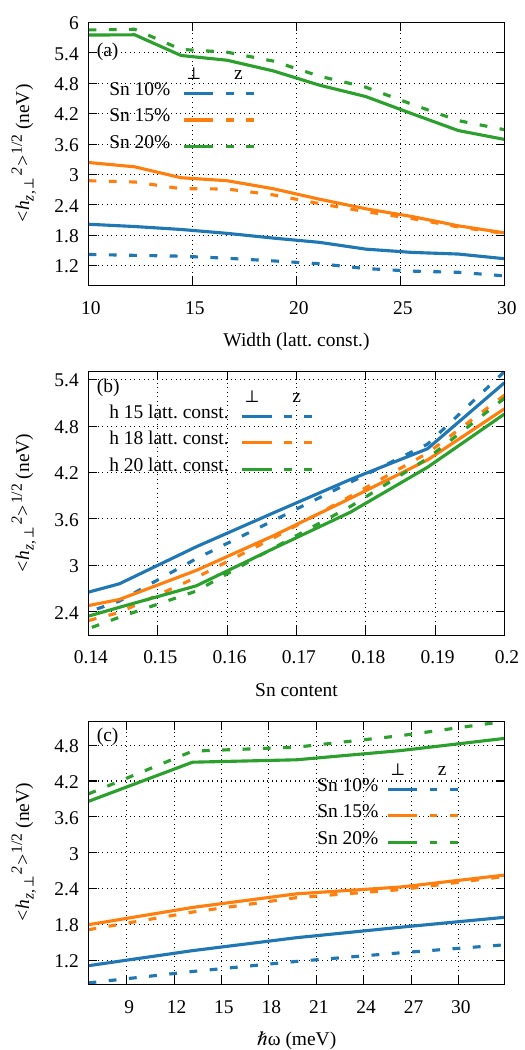}
    \caption{Fluctuations of the Overhauser field represented by rms as a function of:  the QW height (a), the barrier composition (b), and the QD potential characteristic energy $\hbar \omega$ (c). The solid lines denote the averaged transverse components and the dashed lines describe the $z$ component.}
    \label{fig:overhauser}
\end{figure}

We performed comprehensive real-space tight-binding calculations for the hole in an electrically defined QD. Fig.~\ref{fig:en_h}(b) shows the hole ground-state energy as a function of the parameter $\omega$ (for convenience $\hbar \omega$) defining the parabolic in-plane potential, for a fixed QW width of $h = 23$a, where $a$ is the Ge lattice constant.  As expected, the energy increases with $\omega$. In an ideal case of 2D quantum harmonic oscillator, the energy should follow the linear dependence of $\hbar \omega / 2$. However, since we start with a finite computational box, which already introduces the confinement, this dependence is different for small $\omega$. 

Finally, we calculated the ground-state energy dependence on the QW thickness, while the in-plane potential is kept constant ($\hbar \omega = $ 19.8 meV). The results for three barrier compositions $x$ (10\%, 15\%, and 20\% of Sn) are shown in Fig.~\ref{fig:en_h}(c). The energy decreases with increasing $h$, as the confinement gets weaker in the $z$ direction. The most pronounced dependence appears at high Sn content, due to the deepest confining potential well in that scenario.

\subsection{The Overhauser field}

We calculated the Overhauser field fluctuations for the two lowest hole states in the electrically-defined QD as a function of the system parameters. 
We took the magnetic field  $B = 0.1$~T in the $z$ direction (the Faraday orientation).
The results are presented in Figs.~\ref{fig:overhauser}(a-c). As can be seen in Fig.~\ref{fig:overhauser}(a), the values of the transverse and the $z$ field components, decrease with increasing QW width. This is consistent with the approximate relation for the Overhauser field fluctuations~\cite{Fischer2008,Machnikowski2019}
\begin{equation*}
    \expval{h^2_n} \propto \frac{1}{N_\mathrm{part}},
\end{equation*}
where $N_\mathrm{part}$ is the wave function participation number~\cite{Kramer1993}, given by
$$N_\mathrm{part} = \qty( \, \varv \int \abs{\psi(\rr)}^4 \dd[3]{r})^{-1},$$ 
where $\varv$ is the volume of the primitive unit cell. When the QW width increases, the hole wave function spreads over a larger area, which enhances the value of $N_\mathrm{part}$, therefore reducing the strength of the hyperfine coupling.

\begin{table}
    \caption{\label{tab:overhauser} Comparison between rms of the Overhauser field components for various degrees of approximation. All values in neV.}
    \begin{ruledtabular}
        \begin{tabular}{lcccc}
            % & \multicolumn{2}{c}{10\% Sn barrier} & \multicolumn{2}{c}{20\% Sn barrier} \\[1.4pt]
            % \hline \\[-0.5em]
            % & $\expval{h^2_\perp}^{1/2}$ & $\expval{h^2_z}^{1/2}$ & $\expval{h^2_\perp}^{1/2}$ & $\expval{h^2_z}^{1/2}$  \\[1.4pt]
            % \hline \\[-0.5em]
            % Full model &  &  & 5.682  & 5.748  \\[1.1pt]
            % No $d$-shell &  &  & 5.730 & 5.957 \\[1.1pt]
            % No contact term &  &  & 1.440 &  1.931 \\[1.1pt]
            % No contrib. from Sn &  &  & 5.350 & 5.451 \\[1.1pt]

            & \multicolumn{2}{c}{10\% Sn barrier} & \multicolumn{2}{c}{15\% Sn barrier} \\[1.4pt]
            \hline \\[-0.5em]
            & $\expval{h^2_\perp}^{1/2}$ & $\expval{h^2_z}^{1/2}$ & $\expval{h^2_\perp}^{1/2}$ & $\expval{h^2_z}^{1/2}$  \\[1.4pt]
            \hline \\[-0.5em]
            Full model & 2.06 & 1.44 & 3.28  & 2.91  \\[1.1pt]
            No contrib. from Sn & 1.98 & 1.38 & 3.13 & 2.78\\[1.1pt]
            No $d$-shell & 2.13 & 1.62 & 3.34  & 3.10 \\[1.1pt]
            No contact term & 1.58 & 0.978  & 1.58 &  1.32 \\[1.1pt]

        \end{tabular}
    \end{ruledtabular}
\end{table}

The values obtained for the transverse and $z$ components are similar, especially for higher Sn contents in the barrier. This behavior contrasts with the findings for GaAs QDs~\cite{Fischer2008}, and InGaAs/GaAs self-assembled QDs~\cite{Eble2009,Prechtel2016,Machnikowski2019}, where $\expval{h^2_z}$ dominates over $\expval{h^2_\perp}$ for the hole. 
In general, if one assumes that the valence band states are composed only of the $p$-type orbital states, the HH states exhibit $\expval{h^2_\perp}= 0$. In contrast, for the LH states, both Overhauser field components are nonzero and they fulfill the relation $\expval{h^2_\perp} = 4 \expval{h^2_z}$~\cite{Testelin2009}. However, the $d$-type orbital contributions~\cite{Chekhovich2013,Machnikowski2019} and band mixing~\cite{Eble2009,Testelin2009,Machnikowski2019} make the picture more complicated, leading to non-zero transverse components in InGaAs/GaAs QD systems.

To understand the behavior of the field component fluctuations, we performed an analysis of the contributing effects. The first row of Table~\ref{tab:overhauser} presents the full results (like in Fig.~\ref{fig:overhauser}), where we consider the results for $h=10$a and $\hbar \omega = 19.8$~meV for two Sn contents in the barrier: 10\% and 15\%. In both cases, the transverse component is larger, but not in the expected ratio for LH, i.e. $\expval{h^2_\perp}^{1/2} = 2 \expval{h^2_z}^{1/2}$.
One should also note that the hole penetrates the GeSn barrier. Although the values of $\mu_I$ in Ge and Sn are similar (see Table~\ref{tab:hf}), Sn has about twice larger abundance of isotopes carrying the non-zero magnetic moment. To assess the contribution from the Sn atoms,  we excluded them from the simulations for comparison purposes. The results (the 2nd row of Table~\ref{tab:overhauser}) show that the overall impact of the tin atoms is on the order of a few percent, yet growing with the Sn content of the barrier. 

To explore the contributions coming from distinct orbitals, we performed further simulations. First, we artificially ``turned off" the $d$-shell contributions by setting $M_\mathrm{dd} = M_\mathrm{sd} = 0$. The discrepancy compared to the full model is not very large, indicating a moderate contribution of these orbitals to the overall results. We also see that the contribution from the $d$ shell adds destructively to other factors. Finally, we consider the model without the hyperfine contact interaction, which is governed by the $s$-shell (and $s^*$) orbital states. The obtained values are considerably smaller (especially at 15\% Sn), which indicates the importance of the Fermi contact part of the interaction, mediated by the $s$-type admixtures to the LH wave functions. As this effect is isotropic, it tends to equalize the values of $\expval{h^2_\perp}^{1/2}$ and $\expval{h^2_z}^{1/2}$. It can be seen that without the contact term, the obtained transverse to $z$ component ratio gets closer to $2$ (especially for small Sn content). However, it does not reach it, due to the impact of the $d$-shell, and band mixing effect. We note that in the presence of the tensile strain, the latter is affected by the strain-related spin-orbit coupling (see terms proportional to $\bm{k}$ and strain tensor elements weighted by $C_4$ deformation potential in Refs.~\cite{Krzykowski2020,Winkler2003}).

As presented in Fig.~\ref{fig:overhauser}(b), the field components increase with Sn content in the barrier ($x$). This is related to the growing atomic $s$-type contributions in the light-hole wave functions, which is due to the decreasing band gap [cf. Fig.~\ref{fig:en_h}(a)]. For example, the sum of all $s$-type 
admixtures ($s$ and $s^*$ with two spin configurations) to the hole ground state is 1.4\% for Ge$_{0.9}$Sn$_{0.1}$, but 6.4\% for the Ge$_{0.8}$Sn$_{0.2}$ barrier. These numbers are much higher than in the case of InGaAs/GaAs QDs~\cite{Bester2003}, where the contact hyperfine interaction was very small for holes. The dependence observed in Fig.~\ref{fig:overhauser}(b) is further enhanced by two additional factors: Firstly, the larger values of $x$ lead to stronger confinement, which decreases the WF participation number. Secondly, the Sn atoms have a larger abundance (than Ge) of isotopes with non-zero magnetic moment, as discussed above.

Finally, we considered the Overhauser field as a function of $\hbar \omega$ [Fig.~\ref{fig:overhauser}(c)]. This parameter controls the in-plane potential $V_\mathrm{ext}$, hence affecting the WF participation number. However, the obtained dependence is rather weak in the considered range of parameters. 

\section{Conclusions}
\label{sec:concl}

In summary, we have analyzed the spin properties of light-hole states in electrically defined quantum dots within GeSn/Ge/GeSn heterostructures. The  Overhauser field fluctuations were computed using the sp$^3$d$^5$s$^*$ tight-binding model supplemented by density functional theory. Our findings highlight the critical role of conduction-valence band mixing in governing the spin effects. Notably, Overhauser field fluctuations show a strong dependence on the Sn concentration in the barrier, which modulates band mixing. We have demonstrated that in the QD system under study, the hyperfine interaction is dominated to a large extent by the Fermi contact interaction mediated by the $s$-type orbital admixtures.

\acknowledgments

Created using resources provided by Wroclaw Centre for Networking and Supercomputing (\url {http://wcss.pl}).

\appendix

\section{Calculation details}
\label{app:details}

\subsection{Strain calculation}
\label{app:strain}

The lattice mismatch between the Ge and Sn lattice constant inevitably leads to strain. To account for this effect, we utilize Martin's Valence Force Field (VFF) model~\cite{Martin1970,Tanner2019}. The model relies on four parameters $k^\mathrm{(r)}$, $k^\mathrm{(\theta)}$, $k^\mathrm{(rr)}$, $k^\mathrm{(r\theta)}$, which describe bond stretching, bond bending, bond-bond stretching, and bond stretching angle bending, respectively~\cite{Tanner2019}. These parameters are tuned to reproduce the target values of the elastic constants ($C_{11}$, $C_{12}$, $C_{44}$), and the Kleinmann parameter $\zeta_\mathrm{K}$. To this end, we used the analytical formulas (within the covalent crystal approximation) given in Ref.~\cite{Tanner2019}. The model parameters used in the calculations are shown in Table~\ref{tab:elastic}. To match the bowing in the lattice constant for the Ge$_{1-x}$Sn$_x$ alloy~\cite{Polak2017}, we increased the equilibrium distance of Ge--Sn atoms in the calculations of strain.

\begin{table}
    \caption{\label{tab:elastic} The material parameters for strain calculations. The values of $C_{11}$, $C_{12}$, $C_{44}$, and $\zeta_\mathrm{K}$ are taken from Ref.~\cite{Tanner2021}.}
    \begin{ruledtabular}
        \begin{tabular}{lccc}
            & \hspace{25pt} Ge \hspace{25pt} & \hspace{4pt} Sn \hspace{5pt} & \hspace{5pt} GeSn \hspace{5pt} \\
            \hline \\[-0.5em]
            $a$ (\AA) &5.647  &\hspace{2pt}6.479  &6.121$^*$, 6.071  \\[1.1pt]
            $C_{11}$  &132.7  &\hspace{2pt}75.4  &95.1  \\[1.1pt]
            $C_{12}$  &49.8  &\hspace{2pt}35.7  &42.2 \\[1.1pt]
            $C_{44}$  &68.6  &35.1  &46.7 \\[1.1pt]
            $\zeta_\mathrm{K}$  &0.509  &\hspace{2pt}0.631  &0.589  \\[1.1pt]
            \hline\\[-0.5em]
            $k^\mathrm{(r)}$  &6.593  &\hspace{2pt}4.651  &5.119 \\[1.1pt]
            $k^\mathrm{(\theta)}$  &0.487  &\hspace{2pt}0.268  &0.334 \\[1.1pt]
            $k^\mathrm{(rr)}$  &0.266  &\hspace{2pt}0.214  &0.280 \\[1.1pt]
            $k^\mathrm{(r\theta)}$  &0.326  &\hspace{2pt}0.200  &0.270 \\[1.1pt]
            \toprule\\[0.1pt]
            \multicolumn{4}{p{0.97\linewidth}}{\rule{0pt}{1.em} $^*$ Expanded value used only in the strain VFF simulation.} \\
        \end{tabular}
    \end{ruledtabular}
\end{table}

\subsection{The basis}

\label{app:basis}
% We approximate the tight-binding orbitals $f_\alpha(\rr - \bm{R}_i) = \braket{\rr}{\bm{R}_i; \alpha}$ by the hydrogen-like orbitals. The details of the implementation are given in Ref.[]. In present work, the radial-part exponents ($\xi_\mathrm{s}$, $\xi_\mathrm{p}$, and $\xi_\mathrm{d}$) for the orbitals localized at Ge and Sn atoms are extracted from the DFT calculation, as described in the Appendix.  The values of all parameters used in the hyperfine interaction simulations, are summarized in Table.~\ref{tab:hf}

As mentioned in Sec.~\ref{sec:states}, we chose the angular momentum basis $\mathcal{B'} =$ \{$s$, $p_{-1}$, $p_{0}$, $p_{1}$, $d_{-2}$, $d_{-1}$, $d_{0}$, $d_{1}$, $d_{2}$, $s^*$\}. However, the conventional TB Hamiltonian~\cite{Slater1954} $H$ is defined in the $\mathcal{B} =$ \{$s$, $p_{x}$, $p_{y}$, $p_{z}$, $d_{xy}$, $d_{yz}$, $d_{zx}$, $d_{x^2-y^2}$, $d_{3z^2-r^2}$, $s^*$\} basis. Therefore, we perform the transformation
\begin{equation*}
    H' = P^\dagger H P,
\end{equation*}
with
\begin{equation*}
P =
\begin{pmatrix}
P_\mathrm{orb} & 0 \\
0 & P_\mathrm{orb}
\end{pmatrix},
\end{equation*}
where the upper (lower) part of the matrix corresponds to the spin $\uparrow$ ($\downarrow$), and
\begin{equation*}
P^\dagger_\mathrm{orb} =
\begin{pmatrix}
1 & 0 & 0 & 0 & 0 & 0 & 0 & 0 & 0 & 0 \\
0 & \frac{1}{\sqrt{2}} & \frac{i}{\sqrt{2}} & 0 & 0 & 0 & 0 & 0 & 0 & 0 \\
0 & 0 & 0 & 1 & 0 & 0 & 0 & 0 & 0 & 0 \\
0 & -\frac{1}{\sqrt{2}} & \frac{i}{\sqrt{2}} & 0 & 0 & 0 & 0 & 0 & 0 & 0 \\
0 & 0 & 0 & 0 & \frac{i}{\sqrt{2}} & 0 & 0 & \frac{1}{\sqrt{2}} & 0 & 0 \\
0 & 0 & 0 & 0 & 0 & \frac{i}{\sqrt{2}} & \frac{1}{\sqrt{2}} & 0 & 0 & 0 \\
0 & 0 & 0 & 0 & 0 & 0 & 1 & 0 & 0 & 0 \\
0 & 0 & 0 & 0 & 0 & \frac{i}{\sqrt{2}} & -\frac{1}{\sqrt{2}} & 0 & 0 & 0 \\
0 & 0 & 0 & 0 & -\frac{i}{\sqrt{2}} & 0 & 0 & \frac{1}{\sqrt{2}} & 0 & 0 \\
0 & 0 & 0 & 0 & 0 & 0 & 0 & 0 & 0 & 1
\end{pmatrix}.
\end{equation*}

The hole states resulting from diagonalization of the Hamiltonian Eq.~\ref{eq:ham} are expressed as linear combinations of the atomic orbitals (in the $\mathcal{B'}$ basis)
\begin{equation*}
    \ket{\psi_\mu} = \sum^{N_\mathrm{A}}_{i,\alpha} w^\mathrm{(\mu)}_{i,\alpha} \ket{\bm{R}_i; \alpha},
\end{equation*}
where the index $i$ goes over the atomic sites, $w^\mathrm{(\mu)}_{i,\alpha}$ are complex coefficients, and $\ket{\bm{R}_i; \alpha}$ is atomic orbital $\alpha$ localized at the atomic site of position $\bm{R}_i$. The wave functions for these orbitals can be written as
\begin{align*}
    f_\alpha(\rr) = \mathcal{R}_\alpha(r) Y^{l_\alpha}_{m_\alpha}(\theta,\varphi),
\end{align*}
where $\mathcal{R}_\alpha(r)$ is the radial function, $l_\alpha$ is the azimuthal quantum number, and $m_\alpha$ is the magnetic quantum number.

\subsection{The hyperfine coupling}

\label{app:hf}

\begin{figure}[t]
    \centering
    \includegraphics[width=0.4\textwidth]{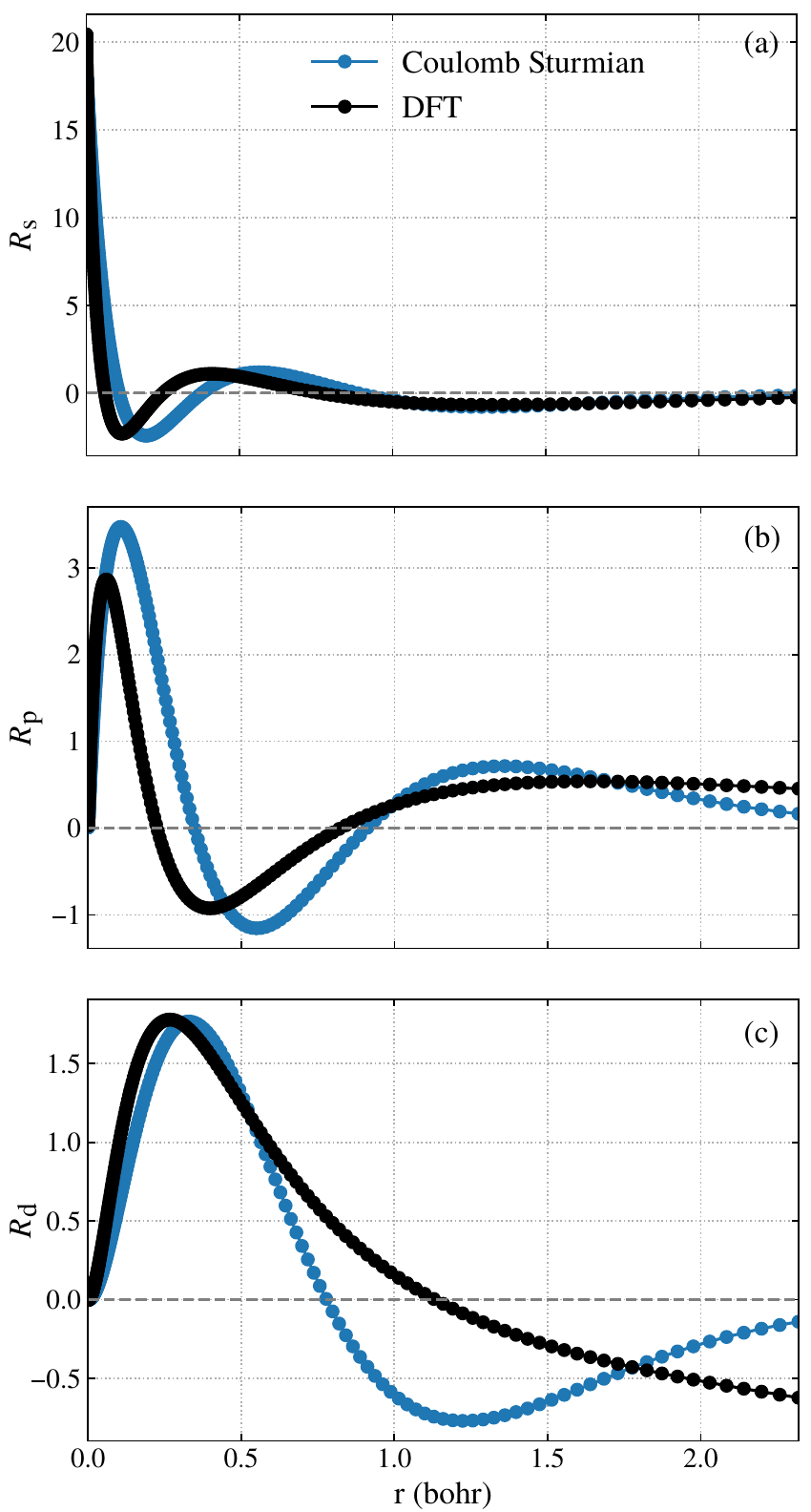}
    \caption{Radial part of the wave function near the atomic core for Ge crystal. (a) Conduction band minimum ($s$ orbital) (b-c) valence band minimum ($p$ and $d$ orbitals, respectively). Blue curves represent a fit to the DFT data with the Sturmian functions 4s, 4p, and 4d with $\xi$ equal to 4.38, 3.97, and 3.85, respectively.}
    \label{fig:wf}
\end{figure}

To calculate the fluctuations of the Overhauser field [Eq.~\ref{eq:hn2}], one needs to determine the matrix elements
\begin{align*}
    A_{n,\mu \nu} &= \mel{\psi_\mu}{A_n(\bm{r} - \bm{R}_i) }{\psi_\nu} \\ & \approx \sum_{i} \sum_{\alpha,\beta} w^{\mathrm{(\mu)}*}_{i,\alpha} 
    w^{\mathrm{(\nu)}}_{i,\beta} \mel{\bm{R}_i; \alpha}{A_n(\bm{r} - \bm{R}_i)}{\bm{R}_i; \beta},
\end{align*}
where we neglected the terms with the orbitals localized at different nuclei than $A_n(\bm{r} - \bm{R}_i)$ is centered on. Therefore, we can simplify the notation to $\mel{\bm{R}_i; \alpha}{A_n(\bm{r} - \bm{R}_i)}{\bm{R}_i; \beta} \equiv  \mel{\alpha}{A_n(\bm{r})}{\beta}$ for a given type of atom.  Owing to the symmetry of the atomic orbitals, it is also convenient to perform further calculations on $\bm{A}$ expressed via spherical vector components
\begin{align*}
    \mathcal{A}_{-1} &= \frac{A_x - i A_y}{\sqrt{2}}, \\
    \mathcal{A}_{0} &= A_z, \\
    \mathcal{A}_{1} &= -\frac{A_x + i A_y}{\sqrt{2}}.
\end{align*}
Their matrix elements are given by~\cite{Machnikowski2019,Gawarecki2025} 
\begin{align*}
    \mel{\alpha}{\mathcal{A}_{q}}{\beta} =& \, \frac{\mu_0 \mu_B}{2 \pi \hbar} \abs{\mathcal{R}_S(0)}^2 \Bigg[ \frac{\sqrt{3} \hbar}{3} \delta_{l 0} \delta_{l' 0} \\ & \times \braket{\frac{1}{2}, 1; s',q }{\frac{1}{2}, 1; \frac{1}{2}, s}  \\ &+ M_{\alpha \beta} \sqrt{l(l+1)} \braket{l, 1; m',q }{l, 1; l, m} \delta_{l l'} \delta_{s s'} \\ 
    & -\sqrt{8 \pi} M_{\alpha \beta}  \sum_{q_1,q_2} \braket{2,1; q_1, q_2}{2, 1; 1, q} \\ & \times G^{m q_1 m'}_{l 2 l'} \frac{\sqrt{3} \hbar}{2} \braket{\frac{1}{2}, 1; s',q_2 }{\frac{1}{2}, 1; \frac{1}{2}, s} \Bigg ], \\
\end{align*}
where we abbreviated $l \equiv l_\alpha$,  $m \equiv m_\alpha$, $s \equiv s_\alpha$;  $l' \equiv l_\beta$,  $m' \equiv m_\beta$,  $s' \equiv s_\beta$; and
\begin{equation*}
    M_{\alpha \beta} = \frac{1}{\abs{\mathcal{R}_S(0)}^2} \int r^2 \mathcal{R}^*_\alpha(r) \frac{1}{r^3} \mathcal{R}_{\beta}(r)  \dd{r}.
\end{equation*}
The elements in the form $\braket{j_1, j_2; m_1, m_2}{j_1, j_2; j, m}$ are the Clebsch-Gordan coefficients.

\subsection{DFT wave function}
\label{app:dft_wf}

As mentioned in the main text, we extracted the AE wave function from the PS wave function using its definition in terms of AE/PS partial waves and projector functions. For this purpose, we used the vaspbandunfolding \cite{vaspbandunfolding} code written by Qijing Zheng. We modified the module that reconstructs the AE wave function on the regular grid to get the angular-momentum-resolved wave function. In particular, we derived the radial part of the AE wave functions on the radial grid and then used it to evaluate the integrals in $M_{\alpha \beta}$ terms. Example functions for the valence band maximum and the conduction band minimum in Ge are presented in Fig.~\ref{fig:wf}. It is known that the former is composed of hybridized $p$ and $d$ orbitals~\cite{Boguslawski1994,Diaz2006,Chekhovich2013}, and the latter is composed of the $s$ orbital. We fitted each function with Coulomb-Sturmian radial functions~\cite{Herbst2019}
\begin{align*}
    \mathcal{R}^\mathrm{(ST)}_{nl} (r) =& N_{nl} \qty(\frac{2 \xi r} {a_\mathrm{B}})^l  \exp(-\frac{\xi r}{a_\mathrm{B}}) L^{2l+1}_{n-l-1}\qty(\frac{2 \xi r}{a_\mathrm{B}}), 
\end{align*}
where $N_{nl}$ is the normalization constant
\begin{equation*}
    N_{nl} = \sqrt{\frac{(2 \xi)^3 (n-l-1)!}{2 n (n+l)! \; a^3_\mathrm{B}}},
\end{equation*}
and $L^{2l+1}_{n-l-1}\qty(x)$ is the generalized Laguerre polynomial.
From Fig. \ref{fig:wf} it is clear that a single Sturmian function is not enough for an accurate description of the wave function for the whole core region. Therefore, using DFT wave function gives more accurate results.

\bibliography{references_KG,references_JZ, references_AM}
	
\end{document}